\documentclass[12pt]{article}

\usepackage{amsfonts}
\usepackage{times}
\usepackage[cmyk]{xcolor}
\usepackage{tikz}
\usepackage{xcolor}
\usepackage{pgfplots}
\usepackage{verbatim}
\usepackage{pgfplotstable}
\usepackage{url}
\usepackage{hyperref}

\pgfplotsset{width=7cm,compat=1.3}

\def\<{\langle}
\def\>{\rangle}
\usepackage{amsmath}
\usepackage{xintexpr}

\title{Classical Simulation of Quantum Supremacy Circuits} 

\author
{Cupjin Huang,$^{1}$ Fang Zhang,$^{2}$ Michael Newman,$^{3}$ Junjie Cai,$^{4}$ \\ Xun Gao,$^{1}$
Zhengxiong Tian,$^{5}$
Junyin Wu,$^{4}$ Haihong Xu,$^{5}$ Huanjun Yu,$^{5}$ \\ Bo Yuan,$^{6}$ Mario Szegedy,$^{1}$ Yaoyun Shi$^{1}$, Jianxin Chen$^{1}$\\
\\
\normalsize{$^{1}$Alibaba Quantum Laboratory,}\\
\normalsize{Alibaba Group USA, Bellevue, WA 98004, USA}\\
\normalsize{$^{2}$Department of Electrical Engineering and Computer Science, }\\
\normalsize{University of Michigan, Ann Arbor, MI 48109, USA}\\
\normalsize{$^{3}$Departments of Physics and Electrical and Computer Engineering,}\\
\normalsize{ Duke University, Durham, NC 27708, USA}\\
\normalsize{$^{4}$Alibaba Cloud Intelligence,}\\
\normalsize{Alibaba Group USA, Bellevue, WA 98004, USA}\\
\normalsize{$^{5}$Alibaba Cloud Intelligence,}\\
\normalsize{Alibaba Group, Hangzhou, Zhejiang 310000, China}\\
\normalsize{$^{6}$Alibaba Infrastructure Service, }\\
\normalsize{Alibaba Group, Hangzhou, Zhejiang 310000, China}\\
}

\date{}

\begin{document} 

\maketitle 

\begin{abstract}
It is believed that random quantum circuits are difficult to simulate classically. These have been used to demonstrate quantum supremacy: the execution of a computational task on a quantum computer that is infeasible for any classical computer.  The task underlying the assertion of quantum supremacy by Arute {\em et al.} ({\em Nature}, {\bf 574}, 505--510 (2019)) was initially estimated to require Summit, the world's most powerful supercomputer today, approximately 10,000 years.  The same task was performed on the Sycamore quantum processor in only 200 seconds.

In this work, we present a tensor network-based classical simulation algorithm.  Using a Summit-comparable cluster, we estimate that our simulator can perform this task in less than 20 days. On moderately-sized instances, we reduce the runtime from years to minutes, running several times faster than Sycamore itself. These estimates are based on explicit simulations of parallel subtasks, and leave no room for hidden costs. The simulator's key ingredient is identifying and optimizing the ``stem'' of the computation: a sequence of pairwise tensor contractions that dominates the computational cost.  This orders-of-magnitude reduction in classical simulation time, together with proposals for further significant improvements, indicates that achieving quantum supremacy may require a period of continuing quantum hardware developments without an unequivocal first demonstration.
\end{abstract}
  
\section{Introduction}
Quantum computers offer the promise of exponential speedups over classical computers.  Consequently, as quantum technologies grow, there will be an inevitable crossing point after which nascent quantum processors will overtake behemoth classical computing systems on specialized tasks.  The term ``quantum supremacy'' was coined to describe this watershed moment \cite{preskill2012quantum}.

Sampling from random quantum circuits is ideal for this purpose. Although it may not have practical value, it is straightforward to perform on a quantum processor: execute a random sequence of quantum gates and then output the measurement result of each qubit. The central challenge is in building a quantum processor of sufficient size and accuracy so that sampling these outcomes becomes infeasible for a classical computer. The hardness of such tasks has been well-studied \cite{aaronson2017complexity, BIS+18, bouland2019complexity, aaronson2019classical,zlokapa2020boundaries}, and is more broadly founded on the expectation that classically simulating general quantum circuits takes a time that grows exponentially with the circuit size. 

In \cite{AAB+19}, the authors sample from a family of random quantum circuits run on the $53$-qubit Sycamore device at increasing depth, measured in terms of ``cycles''.  As quantum gates are noisy, the quantum device samples from a noisy version of the ideal distribution; however, with a sufficient number of samples, the signal can be reliably recovered. Sycamore's performance is a remarkable engineering achievement, and quantum supremacy has been widely discussed. However, it is based on the authors' own best classical runtime estimates. As they themselves point out, simulators will continue to improve, and so these estimates may fall short of the true potential of classical computers \cite{AAB+19}.
  
\subsection{Results}
In this work, we push back against the classical simulation barrier.  We introduce and benchmark a powerful new tensor network-based simulator that dramatically reduces the computational cost of sampling from random quantum circuits. On 53-qubit random quantum circuits with 20 cycles, we estimate that it can sample a near-perfect output in minutes. This reduces the runtime of collecting one million samples from a $20$-cycle circuit adjusted for a $0.2\%$ fidelity from $10,000$ years to less than $20$ days.  On $14$-cycle circuits, we reduce the runtime of collecting three million samples with $1\%$ fidelity from $1.1$ years to $264$ seconds, twice as fast as the total execution time on Sycamore \cite{AAB+19, VBN+19}.  These estimates were obtained by dividing the embarrassingly parallelizable problem into nearly-identical subtasks that run in parallel. These subtasks were then explicitly benchmarked on Summit-comparable nodes within Alibaba Cloud, yielding a precise estimate while avoiding resource-intensive full-scale simulation.

\subsection{Technical contributions and comparisons}
In addition to developing and conglomerating a number of technical optimizations, we achieve these runtimes by introducing a powerful new method for simulating quantum circuits: \emph{stem optimization}.  The key insight is that the random circuits in \cite{AAB+19}, which are subject to the $2$D nearest-neighbor constraints of the device, are highly regular tensor networks.  This regularity manifests at the level of tensor network contraction as a single \emph{stem}: a path of contractions that dominates the overall cost of the computation.  By identifying and optimizing this stem, we are able to increase the efficiency of our simulator to $1,000 \times$ the efficiency of the state-of-the-art simulator introduced in \cite{GK20}, and to approximately $200,000\times$ the efficiency of the original estimate in \cite{AAB+19}. 

\begin{figure}[htb!]
\centering
\includegraphics[width=\textwidth]{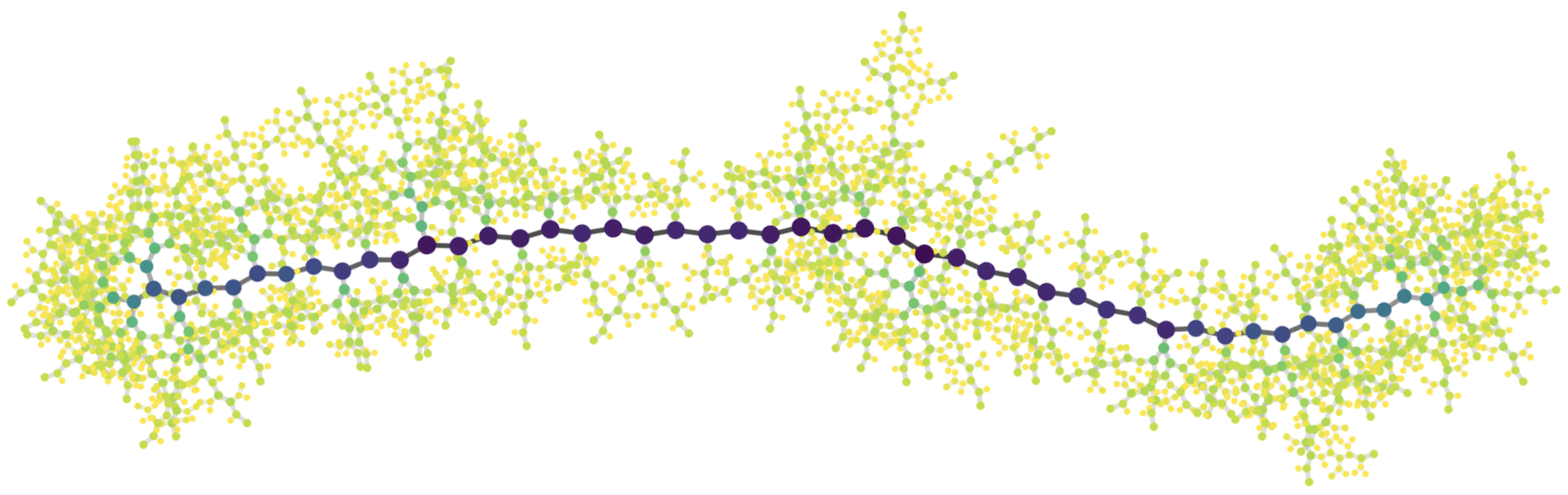}
\caption{Tensor network contraction of a quantum circuit from the random circuit family \cite{AAB+19}, visualized as a binary contraction tree.  Each node in the tree represents a step in the contraction.  Larger, darker nodes indicate more expensive steps.  The central stem dominates the overall contraction cost.}
\label{fig:stembranch}
\end{figure}

Several  challenges  have  been  levied  against  quantum  supremacy.  The  most  relevant  to  this work is not an explicit simulator, but rather a proposal by Pednault et al. \cite{PGN+19} to leverage the immense secondary storage on Summit to perform full state-vector updates \cite{HS17, PGN+17}. It estimates that the quantum supremacy task at 20 cycles can be accomplished in 2.5 days.  There are two main drawbacks to this work.  First, although this simulation strategy scales linearly in depth, it runs into a hard memory limit with even a small increase in the number of qubits.  Second, the estimate is based on optimistic assumptions that are difficult to judge without real device-level tests. Even if we make only a small subset of analogous assumptions, we estimate that running qFlex~\cite{VBN+19} on tasks generated by our algorithm would already reduce the runtime to less than two days. This uses its out-of-core tensor contraction capabilities while assuming that the FLOPS efficiency reported in \cite{AAB+19} is preserved; see Section~\ref{subsection:experiments} of the supplementary material for more details.  We reemphasize that these are proposals; it is difficult to judge their feasibility or accuracy without an explicit simulator tested on a Summit-comparable platform, and may lead to orders-of-magnitude differences.
  
Our simulator produces samples extremely quickly, on the order of minutes for each, so that the overall task remains fast when adjusted for the low fidelity of the quantum computer.  At its core, our tensor network-based algorithm dynamically decomposes the computational task into many smaller subtasks that can be executed in parallel on a cluster. This allows us to trade space for time, avoiding the hard memory limits of full state-vector update simulators \cite{PGN+17,PGN+19}.  Consequently, our simulator will not be overwhelmed by adding a small number of qubits \cite{ZHN+19}. Furthermore, we can accurately estimate the total runtime of our approach using explicit simulations. All of the subtasks have the same tensor network structure, and so their individual runtimes are nearly identical. In addition, each is completed without any communication between nodes. There is indeed some communication when distributing the subtasks and collecting their results, but it is negligibly small. This was verified by a large-scale computational experiment conducted using 127,512 cores of the Elastic Computing Services at Alibaba Cloud; see Section~\ref{subsection:experiments} in the supplementary material.
  
As tensor networks are ubiquitous in quantum information science, with applications including benchmarking quantum devices \cite{AAB+19,AAB+19:data}, probing quantum many-body systems \cite{white1992density, vidal2003efficient, vidal2007classical, schollwock2005density}, and decoding quantum error-correcting codes \cite{bravyi2014efficient,ferris2014tensor,chubb2018statistical}, our simulator represents a powerful new tool to aid in the development of quantum technologies. Technical reports \cite{ZHN+19,HSZ+19,HNZ+20} detail the use of earlier iterations of Alibaba Cloud Quantum Development Platform (AC-QDP) for algorithm design and surface code simulation. 

\section{Sycamore Random Quantum Circuits}
\begin{figure}[htb!]
  \centering
  \includegraphics[width=\textwidth]{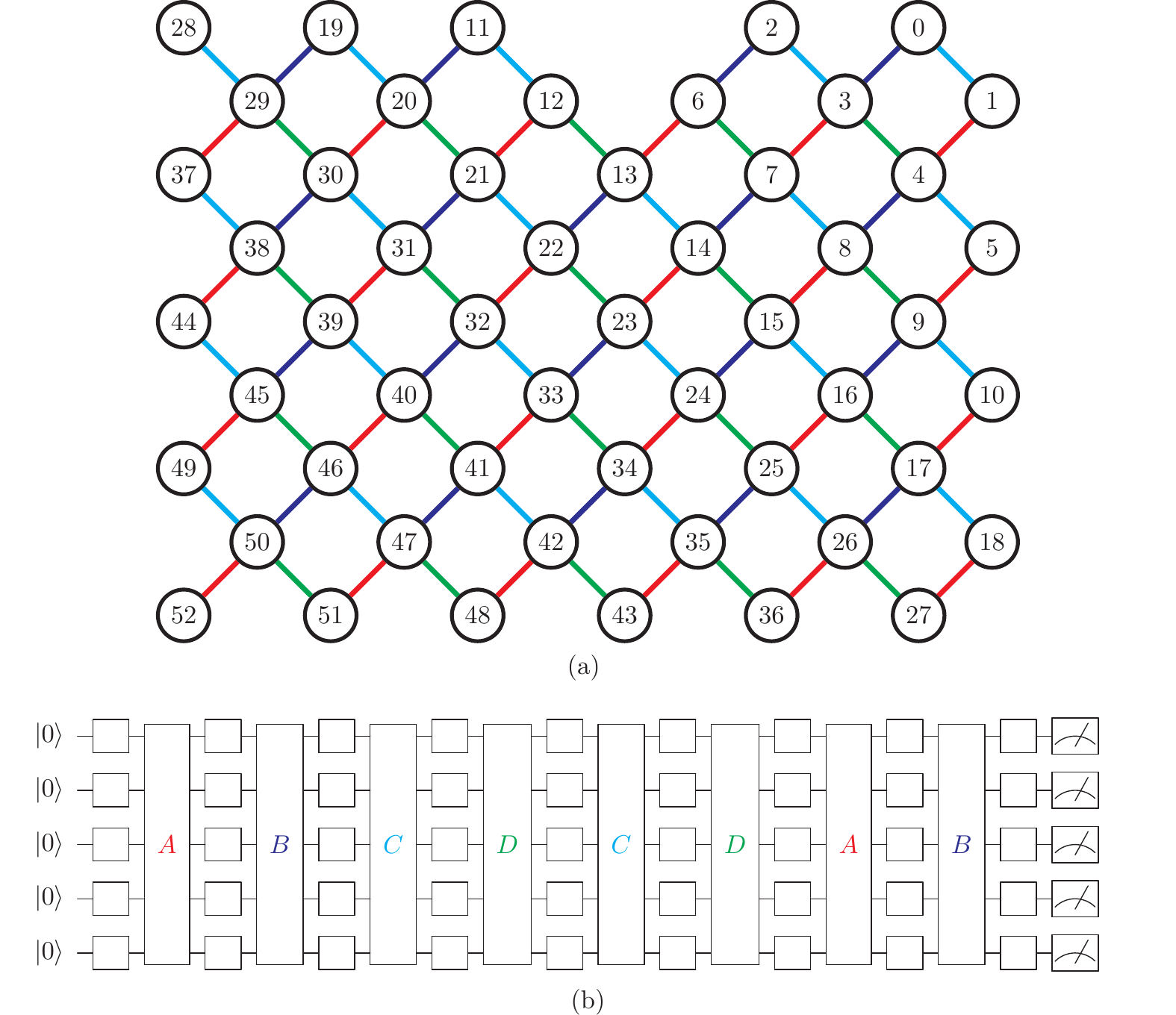}
\caption{Structure of the 53-qubit random quantum circuits simulated in this work. (a) Layout of the 53 qubits and the two-qubit gates between them. Lines of different colors represent two-qubit gates that appear in different layers. (b) Schematic diagram of an 8-cycle circuit. Each cycle includes a layer of random single-qubit gates (empty squares in the diagram) and a layer of two-qubit gates (labeled A, B, C, or D, and colored according to the two-qubit gates in (a)). For longer circuits, the layers repeat in the sequence A, B, C, D, C, D, A, B. Note that there is an extra layer of single-qubit gates preceding measurement.}  \label{fig:circuit}
\end{figure}

In this work, we focus on the random quantum circuits executed on the $53$-qubit Sycamore quantum chip \cite{AAB+19}. Every random circuit is composed of $m$ cycles, each consisting of a single-qubit gate layer and a two-qubit gate layer, and concludes with an additional single-qubit gate layer preceding measurement in the computational basis. In the first single-qubit gate layer, single-qubit gates are chosen for each individual qubit independently and uniformly at random from $\{\sqrt{X},\sqrt{Y}, \sqrt{W}\}$, where $$\sqrt{X}=\frac{1}{\sqrt{2}}\begin{bmatrix}1 & -i \\ -i & 1\end{bmatrix},\hspace{.1cm} \sqrt{Y}=\frac{1}{\sqrt{2}}\begin{bmatrix}1 & -1 \\ 1 & 1\end{bmatrix},\hspace{.1cm}\sqrt{W}=\frac{1}{\sqrt{2}}\begin{bmatrix}1 & -\sqrt{i} \\ \sqrt{-i} & 1\end{bmatrix}.$$ In each successive single-qubit gate layer, single-qubit gates are chosen for each individual qubit uniformly at random from the subset of $\{\sqrt{X},\sqrt{Y}, \sqrt{W}\}$ that excludes the single-qubit gate applied in the previous cycle. In two-qubit gate layers, two-qubit gates are applied according to a specified pairing of qubits in different cycles. There are four different patterns of pairings, depicted in Figure~\ref{fig:circuit}, and we repeat the $8$-cycle pattern A, B, C, D, C, D, A, B. Two-qubit gates are decomposed into four $Z$-rotations determined by the cycle index and $$\textbf{fSim}(\theta,\phi)=\begin{bmatrix} 1 & 0 & 0 & 0 \\ 0 & \cos(\theta) & -i \sin(\theta) & 0 \\ 0 & -i \sin(\theta) & \cos(\theta) & 0 \\ 0 & 0 & 0 & e^{-i\phi}\end{bmatrix},$$ where the parameters $\theta$ and $\phi$ are determined by the qubit pairing.

For noisy quantum devices, a measure is needed to assess the closeness of the output distribution to the ideal Porter-Thomas distribution. The linear cross-entropy benchmarking fidelity (XEB) was used in \cite{AAB+19} for this purpose. It is defined as $2^n \langle p_I(x) \rangle - 1$, where $n$ is the number of qubits, $p_I(x)$ is the probability of $x$ in the ideal distribution, and the expectation is taken over the output distribution. The XEB is $0$ when the output distribution is uniform, and is $1$ when the output distribution is ideal. Simplified quantum circuits run on Sycamore achieved an XEB of approximately $0.2\%$ at 20 cycles \cite{AAB+19}.  It was argued from numerical evidence that the aforementioned random quantum circuits had also achieved an XEB of approximately $0.2\%$. However, simulating these circuits was estimated to be infeasible, and so this could not be directly verified.

\section{Classical Simulation with Tensor Networks}
We view a quantum circuit as a {\em tensor network}.  A tensor network is an attributed multi-hypergraph $H=(V, E)$, where each vertex $v$ is associated to a multi-dimensional array $T_v$ called a tensor, and each hyperedge is associated to a shared index among all the tensors it connects. A \emph{path} is an assignment to the hyperedges that yields an entry of each tensor.  The product of these entries is the value of the path, and the value of a tensor network is the sum of the values of all possible paths. When a subset of hyperedges $E_o\subset E$ is \emph{open}, the value of the tensor network is itself a tensor where each entry corresponds to an assignment of the open edges. Its entries are the summation of the values of all paths that agree with the corresponding assignment.

We adopt the tensor network contraction framework proposed in \cite{BIS+18, AAB+19} as the basis for our simulation of random circuit sampling. This framework assumes that the outcome distribution of a random quantum circuit is a randomly permuted Porter-Thomas distribution. Under this assumption, we can perform {\em frugal rejection sampling} on bitstrings by computing the corresponding amplitudes \cite{MFI+18}. When the batch size of bitstrings is sufficiently large (chosen in our case to be $64$), then with high probability, at least one outcome among the batch will be accepted. We can choose the batch to be a state vector on $6$ qubits while randomly post-selecting the remaining $47$ qubits. In this case, the aggregated result of the amplitudes can be expressed as an open tensor network. This translates the task of sampling from random quantum circuits to the task of contracting a tensor network. Following the observation in \cite{VBN+19}, simultaneously evaluating $64$ amplitudes as an open tensor network yields more efficient sampling, as it is not significantly more costly than evaluating a single amplitude as a closed tensor network. By comparison, performing rejection sampling on single amplitude calculations leads to an approximately 10$\times$ overhead. For random circuits with $m=12,14,20$ cycles, we choose the qubits $(0,1,2,3,4,5)$ in the upper-right corner, and for $m=16, 18$ cycles, we choose the qubits $(10,17,26,36,27,18)$ in the lower-right corner. These choices minimize the overhead introduced by simultaneously evaluating each batch of amplitudes.

There are two approaches for performing tensor network contraction, namely the Feynman method and the Schr\"odinger method.
\begin{description}
  \item[The Feynman method] The Feynman method computes the value of each path separately, and sums over all possible paths. This takes a polynomial amount of space to store the current result and path, but takes a number of steps that scales exponentially with the number of hyperedges. Such high time complexity renders this approach infeasible for even small tensor networks consisting of only $50$ to $100$ hyperedges.
  \item[The Schr\"odinger method] The Schr\"odinger method performs sequential pairwise contractions. At each step, two vertices in the hypergraph are chosen. The corresponding tensors are contracted according to their shared indices, and the two-vertex subgraph is replaced by a single vertex representing the newly formed tensor. This process is repeated until only one vertex is left. Up to transposition, this tensor is the final result and does not depend on the ordering of vertices contracted at each step. Although we can find a contraction order that nearly minimizes the total time complexity \cite{markov2008simulating}, the space complexity scales exponentially with the \emph{contraction width} of the hypergraph \cite{o2019parameterization}. This usually exceeds the accessible memory of a single computational device when simulating intermediate-size quantum circuits.
\end{description}

We use a hybrid method to find an acceptable tradeoff between space and time. We first choose a small subset of indices. For each assignment of these indices, the computational subtask is itself a tensor network contraction, where each sub-tensor network corresponds to the full tensor network with the fixed hyperedges removed. Contracting these sub-tensor networks is perfectly parallellizable, and the space complexity of each subtask is significantly smaller than that of the full task. The hybrid method allows one to trade between space and time, but it requires cleverly choosing an efficient contraction scheme, which includes a good subset of indices as well as an efficient contraction sequence for each subtask.

Determining an optimal contraction scheme is itself a computationally hard problem. However, extensive work has been devoted to finding good contraction schemes, including the introduction of hyperedges \cite{BIS+18}, dynamic edge-slicing \cite{CZH+18}, open tensor network contraction \cite{VBN+19,HNZ+20}, and hypergraph-decomposition-based contraction \cite{GK20}.  Our simulator is a culmination of advances, beginning with \cite{CZH+18} and extended in \cite{ZHN+19,HSZ+19,HNZ+20}. By incorporating a series of newly developed algorithmic ideas and polishing existing ones, we are able to find a good contraction scheme in a reasonable amount of time. 

\section{Stem Optimization}
The main insight of our work is not \emph{how} to optimize the contraction sequence but \emph{where} to optimize it. We observe a generic structure appearing in this family of random quantum circuits: a computationally-intensive \emph{stem} that emerges in a typical contraction tree associated to a circuit's tensor network. A \emph{contraction tree} is a binary tree whose leaf nodes represent initial tensors and whose internal nodes represent intermediate tensors formed from the pairwise contraction of their children.  We can associate to each internal node the time and space complexity of its corresponding pairwise contraction step. 

We observe that a typical contraction tree consists of a single path of expensive nodes which we call the stem, along with small clusters of inexpensive nodes which we call branches, as in Figure~\ref{fig:stembranch}. Almost all of the computation happens along the stem, where a single big tensor absorbs small tensors representing the results from each branch sequentially. Moreover, the length of the stem is usually small compared to the total number of nodes in the contraction tree.

By focusing our attention on the stem, we are able to achieve tremendous speedups in the simulation of random quantum circuits.  The techniques we use to realize these speedups, which we call \emph{stem optimization}, include the following (see Section~\ref{subsection:algorithms} of the supplementary material for more details).

\begin{description}
  \item[Hypergraph partitioning] Introduced in \cite{GK20}, hypergraph partitioning can find a good contraction tree in a top-down manner. It divides the vertices of a hypergraph into several components so that the size of each component is neither too small nor too big, while minimizing the number of interconnecting hyperedges. It first contracts each component to a single vertex, and then contracts the remaining nodes together. The sub-contraction tree associated to each component can be computed recursively while the top-layer of the contraction tree that merges the components can be found by brute force. This brute force search will have low cost because of the small number of interconnecting edges. 
   
  Unlike \cite{GK20}, we first multi-partition at the top level to find one or two major components containing the stem, and then recursively bipartition the stem components to strip off the branches one by one.  A single run of recursive hypergraph partitioning yields a candidate contraction tree. Using the contraction cost of the tree as our objective function, we optimize the parameters of this process (such as the imbalance parameters for each layer) over multiple runs of the recursive partitioning.
  \item[Local optimization] After constructing a good initial contraction tree, we perform local optimizations to further reduce the contraction cost. A connected subgraph of the contraction tree is itself a contraction tree, and its internal connections can be rearranged without affecting other parts of the tree. This allows us to arbitrarily select small connected subgraphs of the contraction tree where brute-force optimization is feasible, and optimize over the chosen subgraphs. We focus on optimizing subpaths of the stem, as this is where most of the computation occurs. Figure~\ref{fig:lo} in the supplementary material illustrates local optimization by switching two branches.
  \item[Dynamic slicing] To simultaneously determine a subset of indices to enumerate as well as a good contraction tree for each sub-tensor network, we first find a good contraction tree for the full task and then choose indices based on that tree. In a contraction tree, all nodes with a shared index form a subtree, and enumerating over that index in the tensor network removes it from the resulting contraction tree. Dynamic slicing was first introduced by us in \cite{CZH+18}, and subsequently adopted in several other simulators \cite{VBN+19,ZHN+19,HNZ+20,GK20,schutski2020simple}. 
  
  Since the stem is the bottleneck for both time and space, the hyperedges whose subtrees intersect most often with the stem are chosen. We interleave selecting hyperedges for enumeration with local optimizations. This process is repeated until the space complexity of each subtask fits into the accessible memory of our computational devices.
\end{description}

\begin{figure}[htb!]
  \centering
  \includegraphics[width=\textwidth]{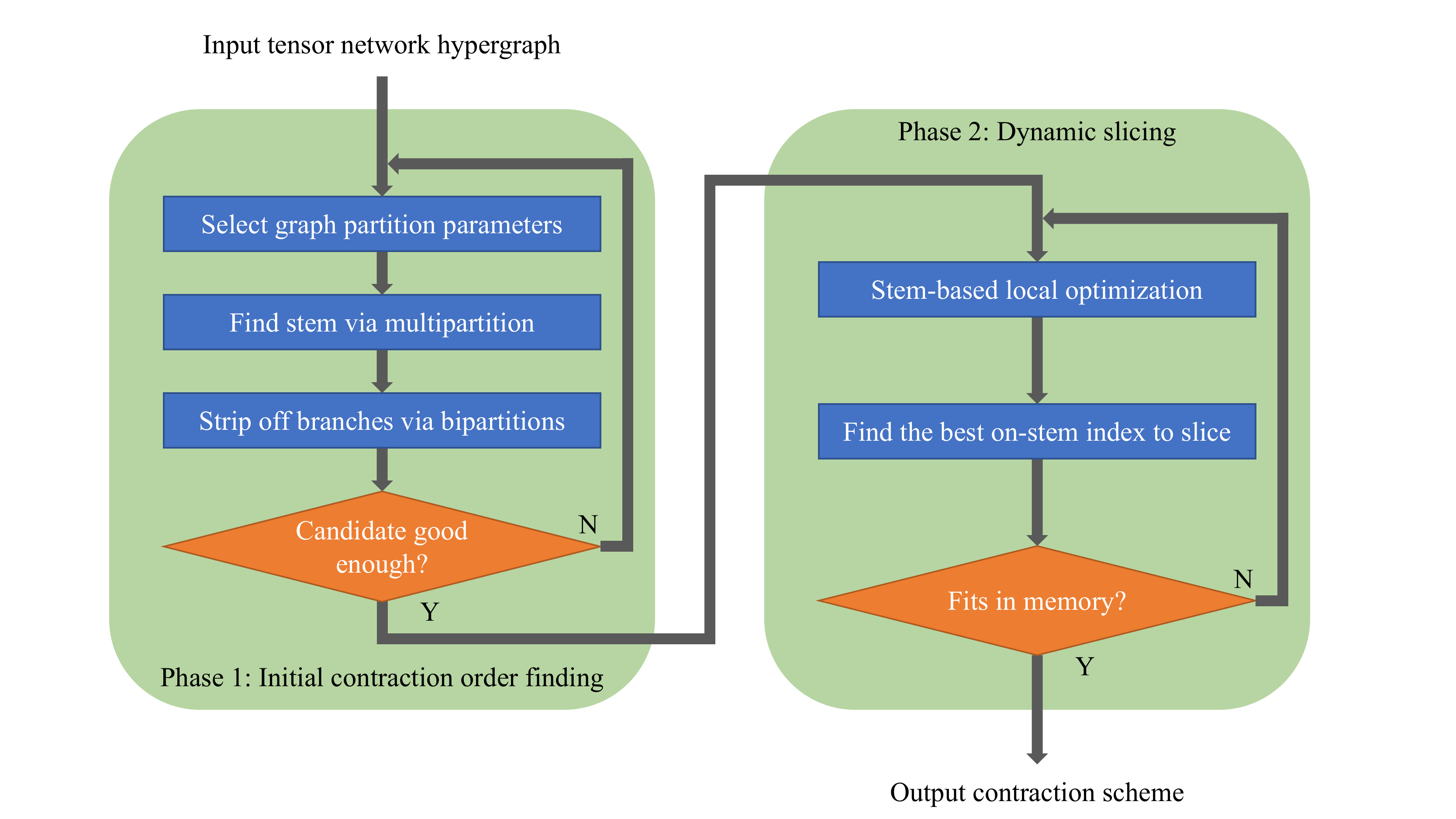}
  \caption{A flowchart describing the algorithm framework; see Section~\ref{subsection:algorithms} for more details. In Phase 1, optimization over hypergraph partition parameters, together with iterative hypergraph decompositions, yields a good unsliced contraction order. In Phase 2, we apply local optimizations to the candidate order along with iterative dynamic slicing to reduce the space complexity.  This process is repeated until a good contraction scheme is found.}
  \label{fig:flowchart}
\end{figure}

\section{Complexity of Random Quantum Circuit Sampling}
We benchmark our simulator, which forms the computational engine of AC-QDP, on random quantum circuits. The circuit files we used are drawn from the data supplement to \cite{AAB+19} and are available at a public Dryad repository \cite{AAB+19:data}. We chose the latest version (Jan.23) that includes circuit files with $12$, $14$, $16$, $18$, and $20$ cycles.

In the preprocessing step for generating a contraction order, we allow for $50$ iterations of \texttt{CMA-ES} for parameter optimization and choose the number of local optimization iterations before, between, and after slicing to be $20$, $20$, and $50$, respectively. The generated contraction order is then distributed to the computational nodes. We run each circuit $5$ times and choose the best contraction order. Figure~\ref{fig:runs} in the supplementary material shows the contraction cost for each run.  Detailed information about our cluster architecture can also be found in Section~\ref{subsection:experiments}.

AC-QDP achieves an exceptionally low contraction cost, up to $10^6$ times lower than qFlex\cite{VBN+19} and up to $10^3$ times lower than Cotengra\cite{GK20}. However, the FLOPS efficiency of AC-QDP is also significantly lower than Cotengra and qFlex. This is likely due to the involvement of many general matrix-matrix products (GEMM) with small-sized matrices during the computation. Overall, AC-QDP  achieves a speedup of more than five orders of magnitude when compared with the best classical algorithms reported in \cite{AAB+19}, and a speedup of more than two orders of magnitude when compared with other state-of-the-art simulators. 

\begin{figure}
  \centering
  \includegraphics[width=0.82\textwidth]{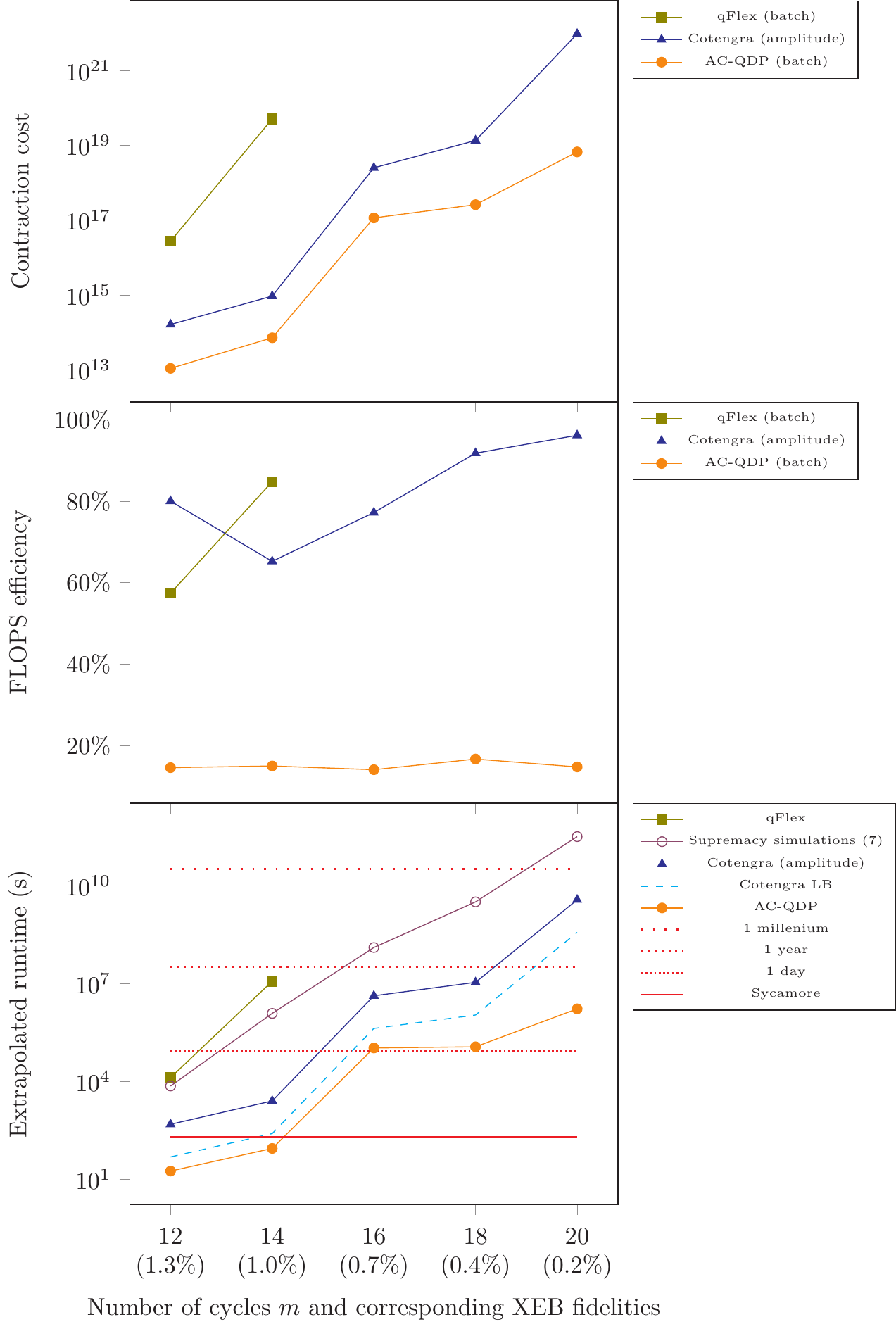}
  \caption{Classical simulation cost of sampling from $m$-cycle random circuits with low XEB fidelities.  Numerical data is reported in Table~\ref{tab:sycamore}.}
  \label{fig:benchmark}
\end{figure}

The contraction cost, FLOPS efficiency, extrapolated runtime, and comparison with other leading simulators are all illustrated in Figure~\ref{fig:benchmark}.  For qFlex and AC-QDP, a batch of amplitudes is computed using open tensor network contraction, while for Cotengra, single amplitudes are computed using closed tensor network contraction.\footnote{Through private communication with one of the authors, we learned that Cotengra has also implemented open tensor network contraction, and that the batch cost may be similar to the single amplitude cost.} In the top subplot, the dark blue curve shows the contraction cost of calculating a single amplitude while the other two curves (qFlex and AC-QDP) show the contraction cost of calculating a batch of $64$ amplitudes. In the bottom subplot, the dark blue curve shows the extrapolated runtime using frugal rejection sampling with a $10\times$ overhead based on the single-amplitude calculation in Cotengra \cite{MFI+18}, while the light blue curve shows the runtime when assuming that calculating $64$ amplitudes can be made as efficient as calculating a single amplitude. We expect that the estimate will approach this lower bound when using open tensor networks in Cotengra.

\begin{table}[htb!]
  \begin{tabular}{ |p{4cm}||p{1.8cm}|p{1.8cm}|p{1.8cm}|p{1.8cm}|p{1.8cm}|  }
    \hline
    \multicolumn{6}{|c|}{Benchmarking results} \\
    \hline
    \hline
    \# Cycles & 12 & 14 & 16& 18 & 20\\
    \hline
    XEB fidelity & 1.3\% & 1.0\% & 0.7\% & 0.4\% & 0.2\% \\
    \# Amplitudes in batch & 64 & 64 & 64 & 64 & 64 \\
    \hline
    \multicolumn{6}{|c|}{Contraction cost and FLOPS efficiency for one sample} \\
    \hline
    Contraction cost  & $1.09\times 10^{13}$ & $7.16\times 10^{13}$ &$1.15\times 10^{17}$& $2.59\times 10^{17}$ & $6.66\times 10^{18}$\\
    FLOPS efficiency  & $14.56\%$ & $14.97\%$ & $14.05\%$ & $16.67\%$ & $14.74\%$\\
    \# Subtasks & $2^7$ & $2^{10}$ & $2^{20}$& $2^{21}$& $2^{25}$\\
    \hline
    \multicolumn{6}{|c|}{Time for generating one perfect sample} \\
    \hline
    One V100  & 38.02s & 243.71s & 4.82d & 9.18d & 266.80d\\
    Summit  & 0.0014s & 0.0088s & 15.06s & 28.69s & 833.75s\\
    \hline    
    \multicolumn{6}{|c|}{Time for generating one million samples with corresponding XEB fidelity} \\
    \hline
    \# Perfect samples  & 13,000 & 10,000 & 7,000 & 4,000 & 2,000\\
    Summit  & 18s & 88s & 1.2d & 1.3d & 19.3d\\
    \hline
  \end{tabular}
\caption{\label{tab:sycamore}Extrapolated benchmarks for the simulation of random circuit sampling on Sycamore using AC-QDP version 2.0 on an Alibaba Cloud cluster comparable with the Summit supercomputer. For $m=12,14$, we explicitly ran one sample on a single V100.} 
\end{table}

\section{Discussion}
There are still many ways to improve the performance of our simulator, potentially by several orders of magnitude. Certainly, a better contraction order might be found through algorithmic refinements. One could also use truncated gates \cite{GK20}, although their overall benefit depends on the consequent decrease in fidelity.
As our current FLOPS efficiency is low ($\approx 15\%$), significant improvements may also be available solely through engineering. On $20$-cycle random quantum circuits, a rough estimation indicates a runtime lower bound using our approach of approximately 12 hours, assuming no performance loss in edge slicing and 100\% FLOPS efficiency on CUDA cores. In addition, we did not take advantage of Tensor Cores, which can deliver up to 83 TFLOPS in mixed precision on a Tesla V100 GPU \cite{MDL+18}. For these reasons, we expect that further improvements on both the algorithmic and engineering sides can significantly reduce the overall simulation costs we report. Given the ubiquity of tensor networks in quantum information science and the efficiency of our simulator, we believe that it could provide a valuable tool for ushering in the development of quantum information technologies while helping to define the quantum supremacy frontier.

\section{Acknowledgments}
\hspace{17pt}\textbf{Contributions}: C.H. and F.Z. contributed equally to this work. We would like to thank our colleagues from various teams in Alibaba Cloud Intelligence and the Search Technology Division for supporting us in the numerical experiments presented in this paper. We also thank Xiaowei Jiang and his team, Yang Yan and his team, and Jianxian Zhang and his local data center team for helping us with the computing facilities. We thank Jiaji Zhu and his team, and Xin Long and his team for their technical support on massive ECS initialization and GPU optimization. Finally, we are particularly grateful to Johnnie Gray for his helpful comments on an earlier version of this manuscript. 

\textbf{Funding}: Part of the work by M.N. and F.Z. was supported through an internship at Alibaba Group USA. M.N. was otherwise supported by ARO MURI (W911NF-16-1-0349). F.Z. was otherwise supported in part by the US NSF under award 1717523.

\textbf{Data and materials availability}: Contraction order files for each circuit will be provided as ancillary files on arXiv. Alibaba Cloud Quantum Development Platform (AC-QDP) will be made publicly available in a few months.

\clearpage

\bibliography{scibib}
\bibliographystyle{unsrt}

\clearpage

\appendix

\section{Supplementary Material}
\subsection{Tensor network contraction}
In this section, we use $[d]$ to indicate the set $\{1,\dots, d\}$ for $d\in\mathbb{N}_+$. For a list of positive integers $\vec{d} = (d_1,\dots, d_n)$, we denote $[\vec{d}]$ as the cartesian product $[\vec{d}]:=\times_{i=1}^n[d_i]$.
\subsubsection{Tensors and tensor networks}
\paragraph{Tensors} A tensor is a multi-dimensional array of complex numbers $T\in \mathbb{C}^{\times_{i=1}^n d_i}$. The number of dimensions $n$ is called the \emph{rank} of the tensor, and the value of each dimension $\vec{d}=(d_1,\dots, d_n)$ is called the \emph{bond dimension} of the tensor network. For a specific index assignment $a=(a_1,\dots, a_n)\in [\vec{d}]$, we use brackets to represent the indexing: $T[a]\in \mathbb{C}$.

\paragraph{Tensor networks} A tensor network is an attributed multi-hypergraph $H=(V,E)$, where
\begin{enumerate}
  \item there is a subset of hyperedges $E_o$ called \emph{open edges}, and
  \item there is a mapping $d:E\rightarrow \mathbb{N}^+$ from each hyperedge to a positive integer, called the \emph{bond dimension} of the hyperedge, and
  \item for each vertex $v$, there is an ordering $o_v: [\deg(v)] \rightarrow \{e\in E| v\in e\}$ of the hyperedges incident to it. Moreover, there is a tensor $T_v\in \mathbb{C}^{\times_{i=1}^{\deg(v)}d(o_v(i))}$ associated to each vertex $v$.
\end{enumerate}

An assignment of all the hyperedges in the tensor network, i.e.\ an element $a\in[(d(e))_{e\in E}]$, fixes indices of all tensors in the tensor network, and is associated to the product of all the corresponding entries via the Feynman path:
$$F(a):=\prod_{v\in V}T_v[(a_{o_v(1)}, a_{o_v(2)},\dots, a_{o_v(\deg(v))})].$$

Given a tensor network $H=(V,E)$, the \emph{value} associated to the tensor network is a tensor $T_H\in \mathbb{C}^{\times_{e\in E_o}d(e)}$ so that the entry of the assignment $b$ is the summation of all Feynman paths whose index assignments agree with $b$:
$$T_H[b]:=\sum_{a\in [(d(e))_{e\in E}]; \forall e\in E_o, a_e=b_e}F(a).$$
The computational task of solving for the value of $T_H$ given the tensor network $H$ is called \emph{contraction} of the tensor network.

\begin{figure}[htb!]
    \centering
    \includegraphics[trim={3cm 3cm 3cm 3cm}, width=\textwidth]{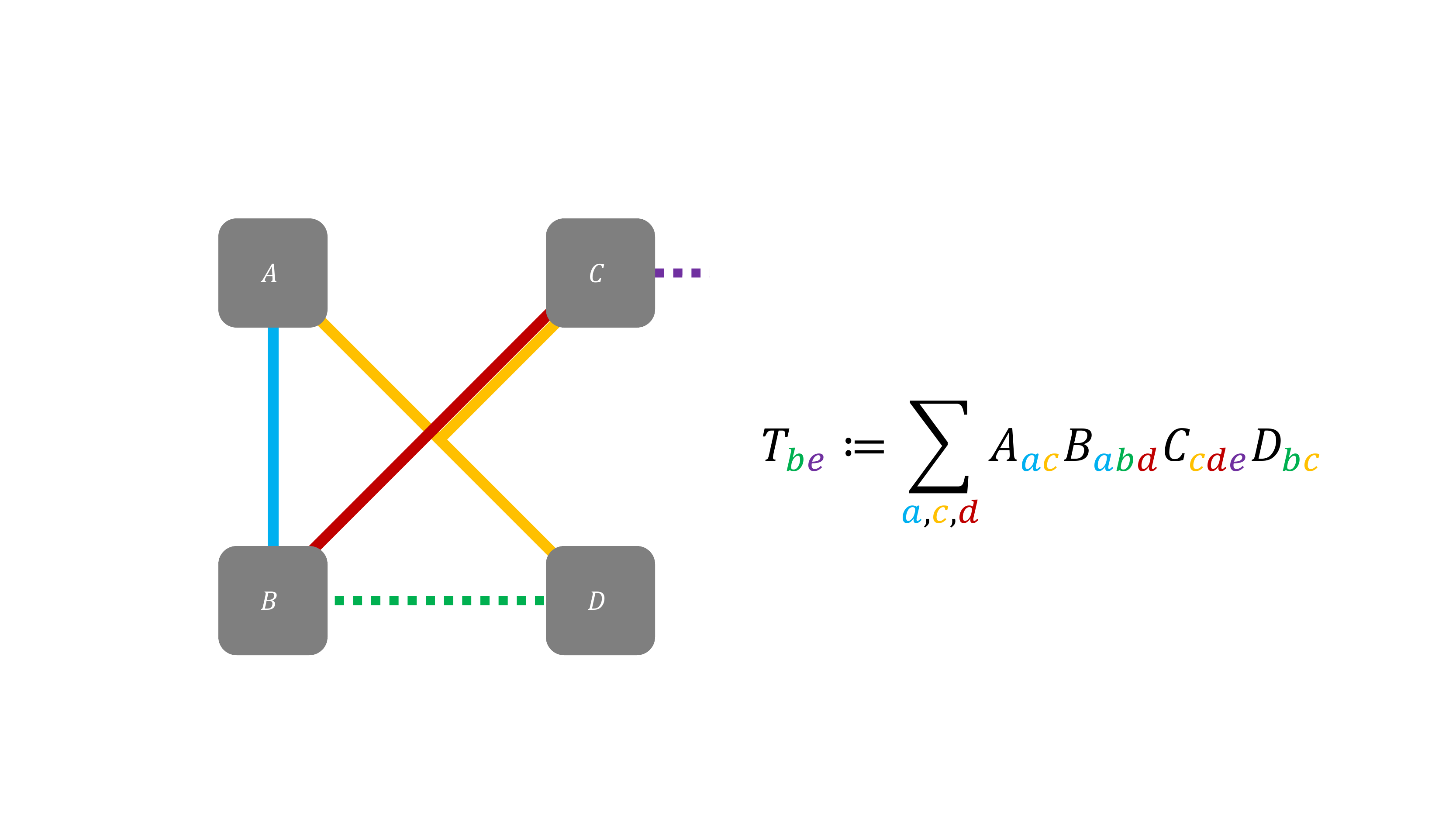}
    \caption{A tensor network consisting of $4$ tensors and $5$ hyperedges.}
    \label{fig:tensor_network}
\end{figure}

For the purpose of classically simulating qubit quantum circuits, it is sufficient to assume that all bond dimensions appearing are 2. Figure~\ref{fig:tensor_network} illustrates a small tensor network consisting of $4$ tensors and $5$ hyperedges.

\subsubsection{Contraction trees and tensor network slicing}
\begin{figure}[htb!]
    \centering
    \includegraphics[width=\textwidth]{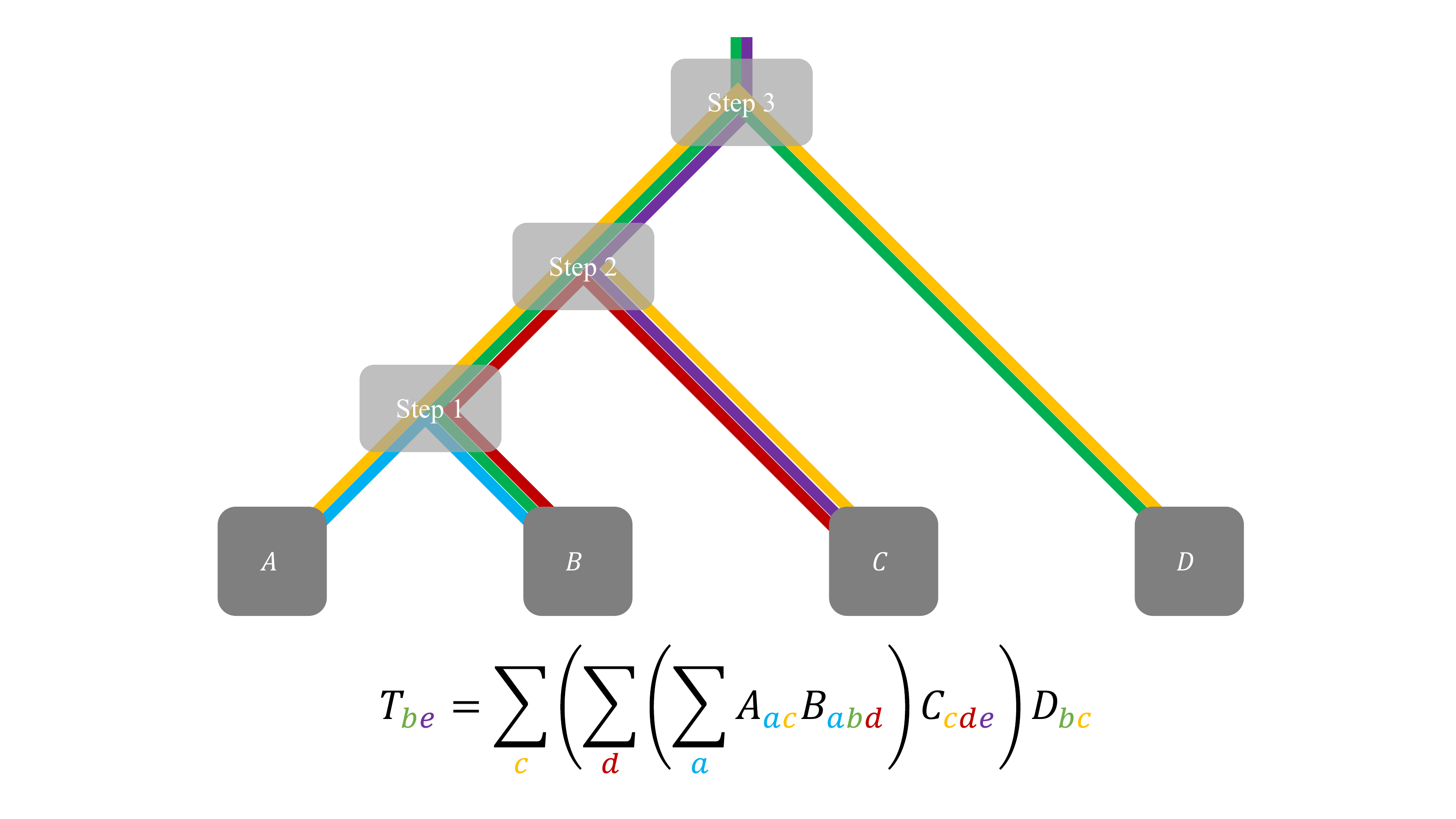}
    \caption{A contraction tree realizing the evaluation of the tensor network shown in Figure~\ref{fig:tensor_network} via step-wise tensor contractions. The number of hyperedges present in a contraction step indicates the time complexity, and the number of outgoing hyperedges indicates the space complexity. The total time complexity is the sum of the time complexities for each step, and the total space complexity is the maximum over all step-wise space complexities. In this particular contraction tree, the total time complexity is $16+16+8=40$ and the total space complexity is $8$.}
    \label{fig:contraction_tree}
\end{figure}

\paragraph{Contraction trees} One common method for performing tensor network contraction is through \emph{sequential pairwise tensor contraction}. At each step, two vertices in the tensor network are chosen. The corresponding subgraph is then replaced by a single vertex preserving all the outgoing connections, with an updated tensor associated to it. Figure~\ref{fig:contraction_tree} depicts a contraction tree that corresponds to the tensor network in Figure~\ref{fig:tensor_network}.

The time and space complexity of one single step of pairwise tensor contraction, without applying Strassen-type accelerations, is defined as follows. Denote the two vertices as $A, B$, let $E_{AB}$ be all of the edges that are either connected to $A$ or $B$, and let $E^*_{AB}\subset E_{AB}$ be the edges in $E_{AB}$ that is an open edge or is connected to other vertices. The stepwise contraction enumerates over all Feynman paths associated with the hyperedges in $E_{AB}$, and sums up those that agrees on $E^*_{AB}$. The resulting time and space complexities for a step are then $2^{|E_{AB}|}$ and $2^{|E^*_{AB}|}$, respectively.  The above process can be repeated until there is only a single vertex left whose associated tensor (up to transposition) is the final result. This process can be abstracted as a rooted binary tree.

The final result of the contraction does not depend on the specific order one chooses to pairwise contract the tensors, but only on the hypergraph structure and the specific values of the input tensors. In contrast, the time (space) complexity, defined as the summation (maximum) of the time (space) complexities of the stepwise contractions, does not depend on the specific values of the input tensors. Rather, it only depends on the order in which they are contracted. This motivates us to ignore the values of the input tensors for now, and focus on the unattributed hypergraph when investigating the contraction trees.

A contraction tree $T=(V\cup V', E_T)$ associated to a tensor network $H=(V,E)$ is a rooted binary tree with root $r$, where each node in $V$ is a leaf node and each node in $V'$ has exactly two children. Each leaf node is associated to the hyperedges it is connected to in $H$, and each non-leaf node $u$ with children $A$ and $B$ is associated to the set $E_u=E_{AB}$. Furthermore, for a non-leaf node $u$ with children $(A,B)$ we denote $E^*_u:=E^*_{AB}$. It can be shown that each hyperedge is present in the minimum connected subgraph of $T$ spanned by the leaf nodes adjacent to it, and the root node for open edges.

The \emph{time complexity} associated to the contraction tree $T$ is defined as
$$tc(T):=\sum_{u\in V'}2^{|E_u|},$$
while the \emph{contraction width} of a contraction tree $T$, which serves as an indicator of the space complexity, is defined as
$$cw(T):=\max_u\in V'|E^*_u|.$$

\begin{figure}[htb!]
    \centering
    \includegraphics[width=\textwidth]{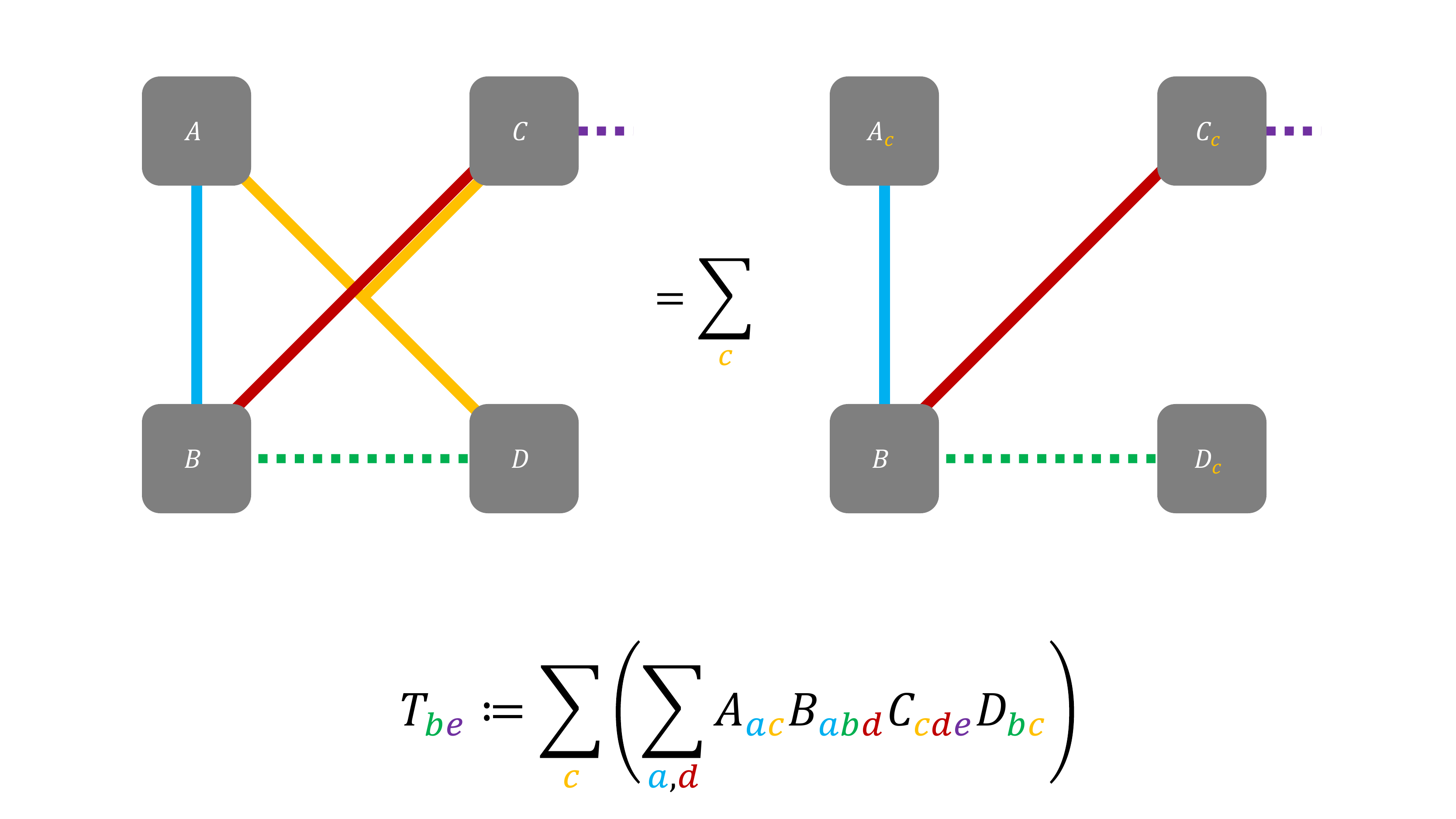}
    \caption{An illustration of index slicing on the tensor network illustrated in Figure~\ref{fig:tensor_network}. Here, the index $c$ is sliced, yielding two sub tensor networks with identical structure, with $c$ assigned $0$ and $1$ respectively.}
    \label{fig:slicing}
\end{figure}

\paragraph{Slicing} Slicing of a tensor network divides a contraction task into subtasks for both embarrassing parallelism and reduction in space complexity. Slicing a tensor network begins by choosing a subset of hyperedges. Each assignment to the indices on those hyperedges induces a sub tensor network. The value of the full tensor network can then be obtained by enumerating over all possible assignments on the chosen hyperedges and then summing the values of the sub tensor networks together, see Figure~\ref{fig:slicing}.

Given the set of sliced hyperedges $E_s$ of a hypergraph $H=(V,E)$, all of the subtasks, though differing in tensor values, are associated to the same sub hypergraph $H_s = (V, E\setminus E_s)$. The sub tensor networks can either be contracted via a contraction tree or sliced recursively. However, the latter can be reduced to the former without loss of generality by regarding the overall slicing as a slicing over the union of all sliced hyperedges. 

A slicing-incorporated tensor network contraction scheme is then a pair $(E_s, T)$ where $E_s$ is the set of sliced hyperedges and $T$ is the contraction tree on the sub hypergraph $(V, E\setminus E_s)$. The time complexity and the contraction width of this contraction scheme are then
$$tc((E_s, T)):=2^{|E_s|}tc(T),$$ $$cw((E_s, T)):=cw(T).$$

Real-world computational devices can usually work for long periods of time, but are restricted by memory. Therefore, the goal of finding a good contraction scheme can be formulated as minimizing the total time complexity subject to the constraint that the space complexity is bounded by a predetermined memory limit. For an Nvidia V100 with $16$ GiB of memory, it is usually sufficient to slice the tensor network so that the contraction width is at most 29, assuming single precision.

\subsection{Optimization methods}
\label{subsection:algorithms}
In the following section, we describe the heuristics we use for obtaining a good contraction scheme. The heuristics are based on our observation that a typical contraction tree consists of a stem of overwhelming cost, and short branches attached to that stem. See Figure~\ref{fig:flowchart} for a flowchart describing the algorithm.

\subsubsection{Stems and branches}
We first focus on finding good contraction trees without slicing indices. Since the total time complexity of a contraction tree is an exponential sum $tc(T)=\sum_{u\in V'}2^{|E_u|}$, it is important to understand how the cost is distributed across a typical contraction tree. We observe that the high-weight nodes in a contraction tree tend to form a path, as illustrated in Figure~\ref{fig:stembranch}. We call this high-weight path the \emph{stem} of the tree. Each node in this path has essentially the same weight, while nodes outside this path typically have significantly smaller weights. Moreover, the nodes with smaller weights tend to form small clusters attached to the stem of the tree we call \emph{branches}. Consequently, we focus on optimizing the stem of the contraction tree by reducing its thickness (the computational cost for each stem node) and its length (the number of stem nodes) in order to minimize the total contraction cost.

\subsubsection{Multi-parameter optimization based on hypergraph decomposition}
\paragraph{Stem finding: hypergraph multipartite decomposition} Inspired by \cite{GK20}, we use hypergraph decompositions to first find the starting point of a stem. A $K$-wise hypergraph decomposition with imbalance parameter $\epsilon$ decomposes the hypergraph into $K$ parts such that the cut across each part is minimized, subject to the constraint that each part contains at most $(1+\epsilon)\lceil\frac{|V|}{K}\rceil$ vertices. It was shown in \cite{GK20} that hypergraph decomposition-based contraction-finding algorithms are more efficient than previously proposed tree decomposition-based algorithms. This is because the cuts are more relevant to the space complexity of the contraction scheme than the treewidth, which is only the maximum time complexity of a single step. In practice, we find that multipartite hypergraph decompositions work best at finding major components containing the stem of the contraction. A hypergraph decomposition typically finds one major component containing the stem, and occasionally it finds two major components each containing part of the stem, in the case that the root of the contraction tree lies in the middle of the stem.

\paragraph{Stem construction: recursive bipartite decomposition} Once we find the major components containing the stem, we can remove the branches one-by-one by recursively using hypergraph bipartitions. Setting an imbalance parameter $\epsilon'$ close to 1 allows us to ``peel off'' one small branch at a time. This process can be repeated until the number of nodes left is fewer than a preset threshold $N$, at which point the stem is believed to have ended.

In our implementation, $N$ is fixed to $25$. The algorithm optimizes over the parameter set $(K, \epsilon, \epsilon')$ in order to find an optimal assignment of the parameters. We use the software packages \texttt{KaHyPar}~\cite{ahss2017alenex, shhmss2016alenex} for hypergraph decomposition and \texttt{CMA-ES}~\cite{hansen2019pycma} for parameter optimization. At the end, a batch of initial contraction trees are obtained from the best parameter set we found, which are then fed to further optimization and slicing routines described below.

\subsubsection{Local optimization}
A connected subgraph of a contraction tree can be regarded as a smaller contraction tree, where its leaf nodes represent the input tensors and the highest level node its output. Moreover, the internal connections of the subgraph are independent of the other parts of the contraction tree, and the total contraction cost is a function of all stepwise costs. Therefore, one can apply local optimization techniques on small subgraphs of an obtained contraction tree.

\begin{figure}[htb!]
  \centering
  \includegraphics[trim={5cm 4cm 4cm 3cm},clip,width=\textwidth]{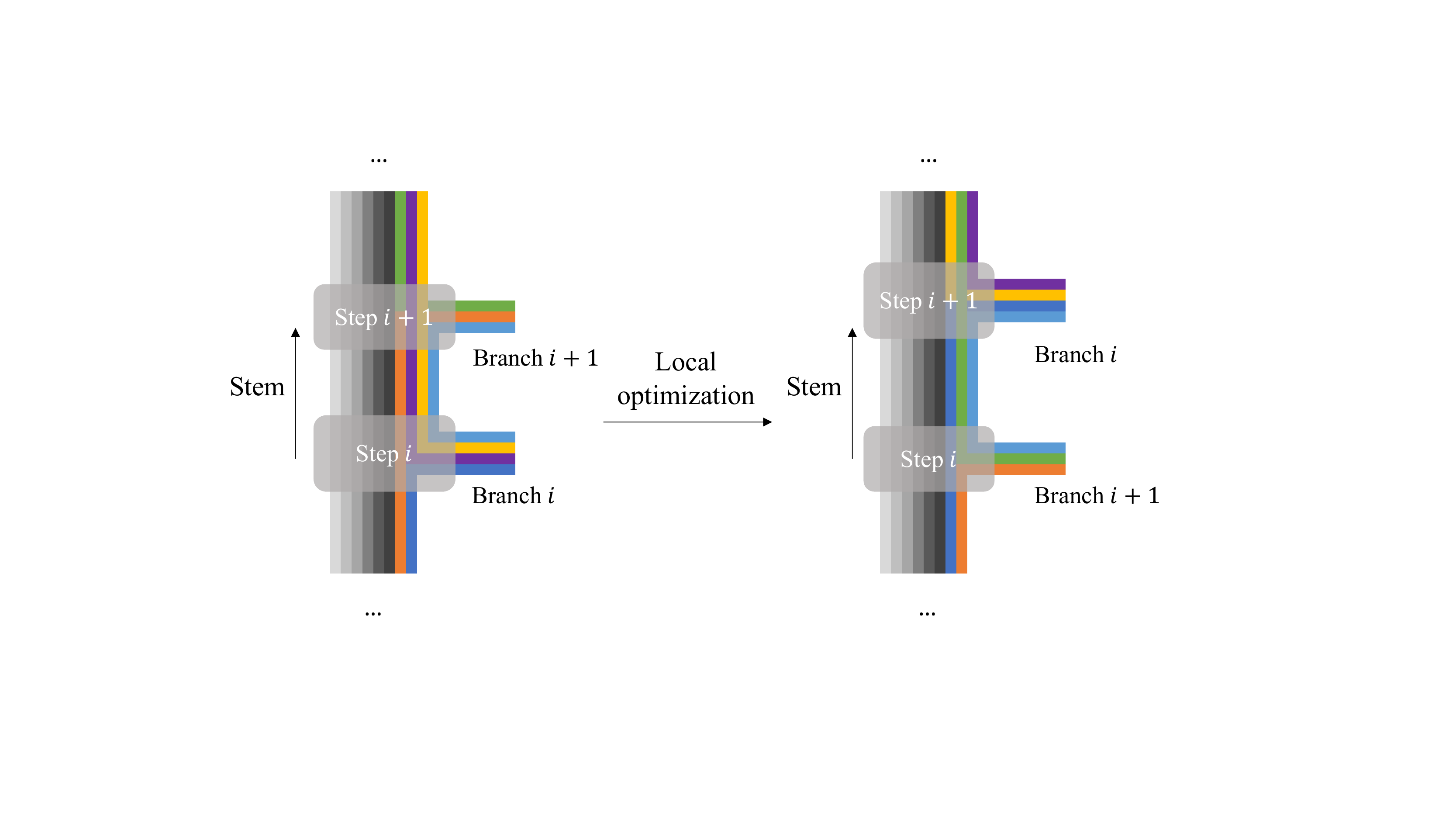}
  \caption{Illustration of stem-based local optimization. By switching the branches $i$ and $i+1$, both the space and time complexity of the contraction are reduced.}
  \label{fig:lo}
\end{figure}

In our implementation, we randomly choose a small connected subgraph with nodes from the stem of the contraction tree, and then find the optimal internal connections to replace the original with brute force. Note that such an algorithm is feasible only when the subgraph is sufficiently small. This process is repeated until no significant improvements are observed, or a fixed number of iterations have passed.

\subsubsection{Dynamic slicing}
When the space complexity of the contraction tree is not small enough to fit into memory, slicing must be done. In our implementation, slicing is performed greedily so that each time a hyperedge is sliced, the total time complexity increases the least. Between two steps of picking sliced indices, local optimization is applied to slightly restructure the contraction tree. This helps to increase the chance that the next sliced hyperedges will not increase the total time complexity significantly. In practice, we observed an $\approx .2\%$ total increase in time complexity for $10$ hyperedges sliced, and about an $\approx 4\times$ increase for $25$ hyperedges sliced.

\subsection{Runtime modification of the contraction scheme}
To better accommodate the capability and limitations of the GPU, we further apply the following runtime-specific modifications.
\paragraph{Pre-computation} The branches of the contraction tree represent insignificant portions of the overall time complexity. However, they involve many small tensors, transmission of which to the GPU would incur significant overhead. Moreover, hyperedges involving branches usually have very little intersection with the sliced hyperedges. This motivates us to pre-compute the branches on a CPU before computing the stems of the individual tasks. The partial results for the branches are shared by all subtasks and only need to be computed once. In practice, this significantly reduces the communication cost between the GPU and the CPU, and helps save a small portion of computational cost.

\begin{figure}[htb!]
  \centering
  \includegraphics[trim={5cm 3cm 3cm 3cm},clip,width=\textwidth]{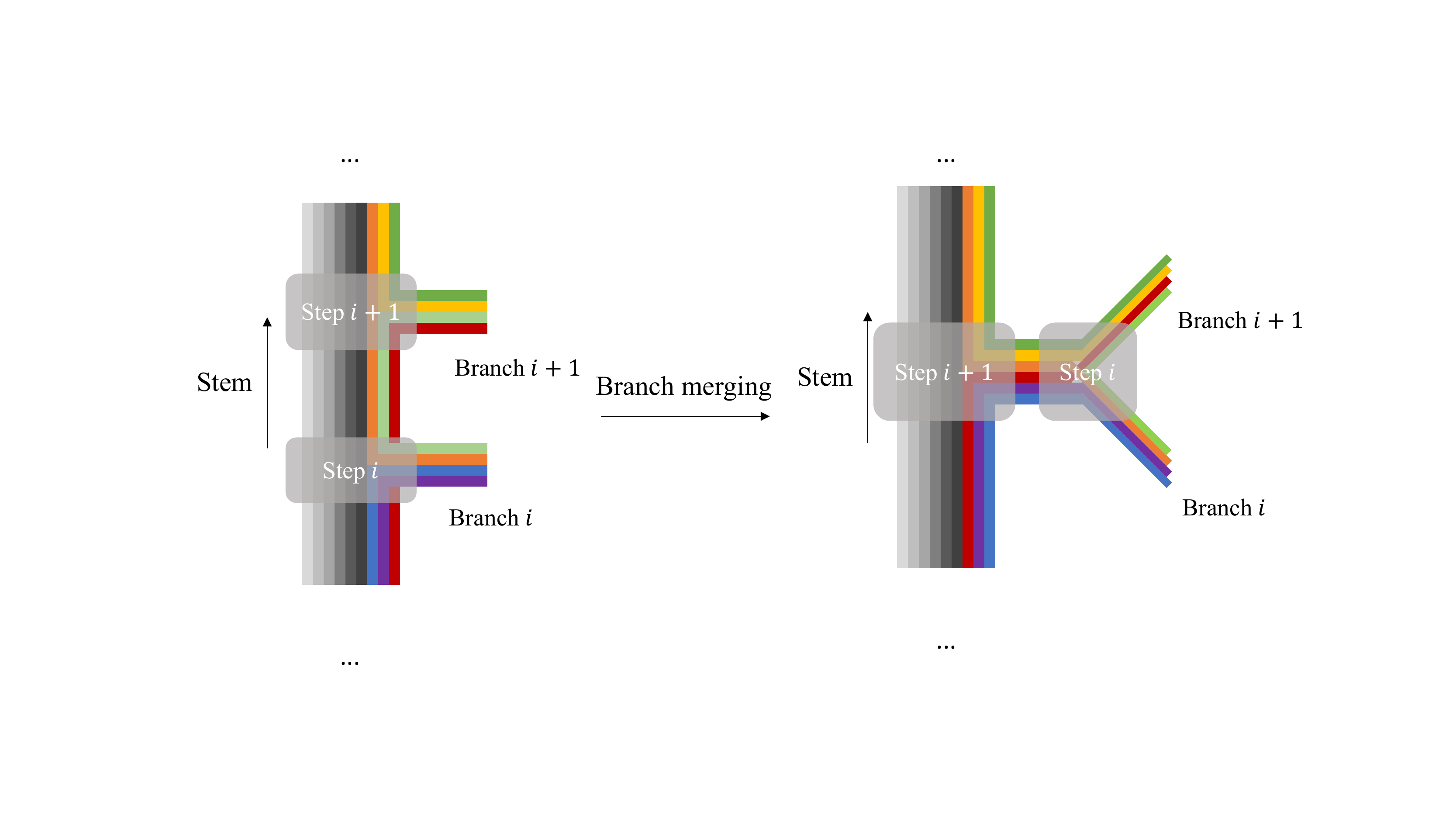}
  \caption{Illustration of a runtime branch merging. GPUs provide significant accelerations for contracting large matrices. By combining branches $i$ and $i+1$ together before merging into the stem, a larger tensor is absorbed all-at-once. While this might slightly increase the number of FLOPs, the patching scheme significantly reduces the actual runtime on modern GPU architectures.}
  \label{fig:patch}
\end{figure}

\paragraph{Branch merging}
The stem computation performed on the GPU is typically a sequential absorption of small branch tensors into the large stem tensor, or two sides of the stem merged together near the root. In either cases, a contraction tree with locally optimal contraction costs typically suffers from a large contrast in dimensions during matrix multiplication. On Nvidia Tesla V100 GPUs, matrix multiplication with dimensions $M\times N$ and $N\times K$ is much more efficient when the dimensions $M, N$ and $K$ are all multiples of $32$.  However, a typical branch tensor is often shaped as $4\times 4$, $8\times 8$, or $16\times 16$. To overcome this, adjacent branches are merged together to form a bigger tensor to contract with the stem, as depicted in Figure~\ref{fig:patch}. This increases the runtime contraction cost, but decrease the actual runtime by making use of the efficient kernel functions of the Nvidia Tesla V100. While this is an ad-hoc solution to the low GPU-efficiency induced by small tensor dimensions, we hope that more systematic ideas could be incorporated to increase the GPU efficiency.

\subsection{Experiments} \label{subsection:experiments}
\subsubsection{Cluster architecture}
We use the Alibaba Cloud clusters to conduct numerical experiments. Though cluster architectures may suffer from higher latency in communication between nodes than a supercomputer, our tensor network-based computation requires little communication. 

The cluster architecture consists of a single node as an agent to split the large tensor network contraction task into many smaller tensor network contraction subtasks. This step is called {\em preprocessing}. Then, the agent node uses the OSS (Object Storage Service) as a data transmission hub to assign different subtasks to different computation nodes. When a computation node is finished with the assigned subtask, it will upload the result to the OSS, and the agent node will repeatedly query the OSS until all the subtask results add up to the desired amplitude. The reported {\em runtime} is the total elapsed time obtained on the agent node except for the preprocessing step. This is because we only need to do preprocessing once on a single node (the agent node), independent of the number of amplitudes we will calculate. Therefore, the preprocessing step is negligible in terms of both runtime and core-hours for sampling tasks that usually involve thousands of amplitudes and many computation nodes.

\begin{figure}[htb!]
  \centering
  \hspace{-1cm}\includegraphics[width=.9\textwidth]{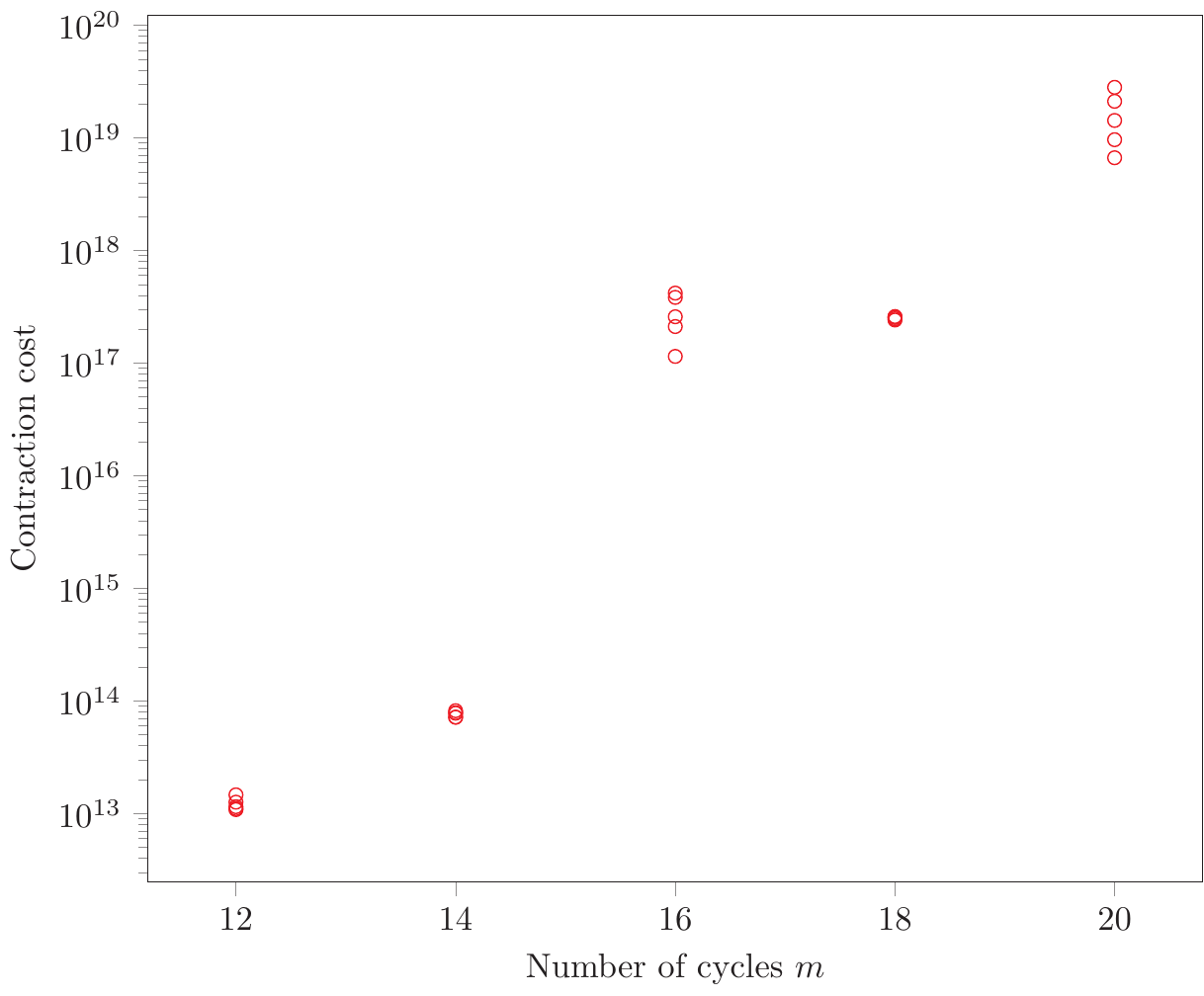}
  \caption{For each length, we run preprocessing for the quantum circuits in \cite{AAB+19:data} five times. In Figure~\ref{fig:benchmark} we report the best contraction order we obtained over all five runs. Here, we report the contraction cost for each run.}
\label{fig:runs}
\end{figure}

\begin{figure}[htb!]
  \centering
  \includegraphics[width=\textwidth]{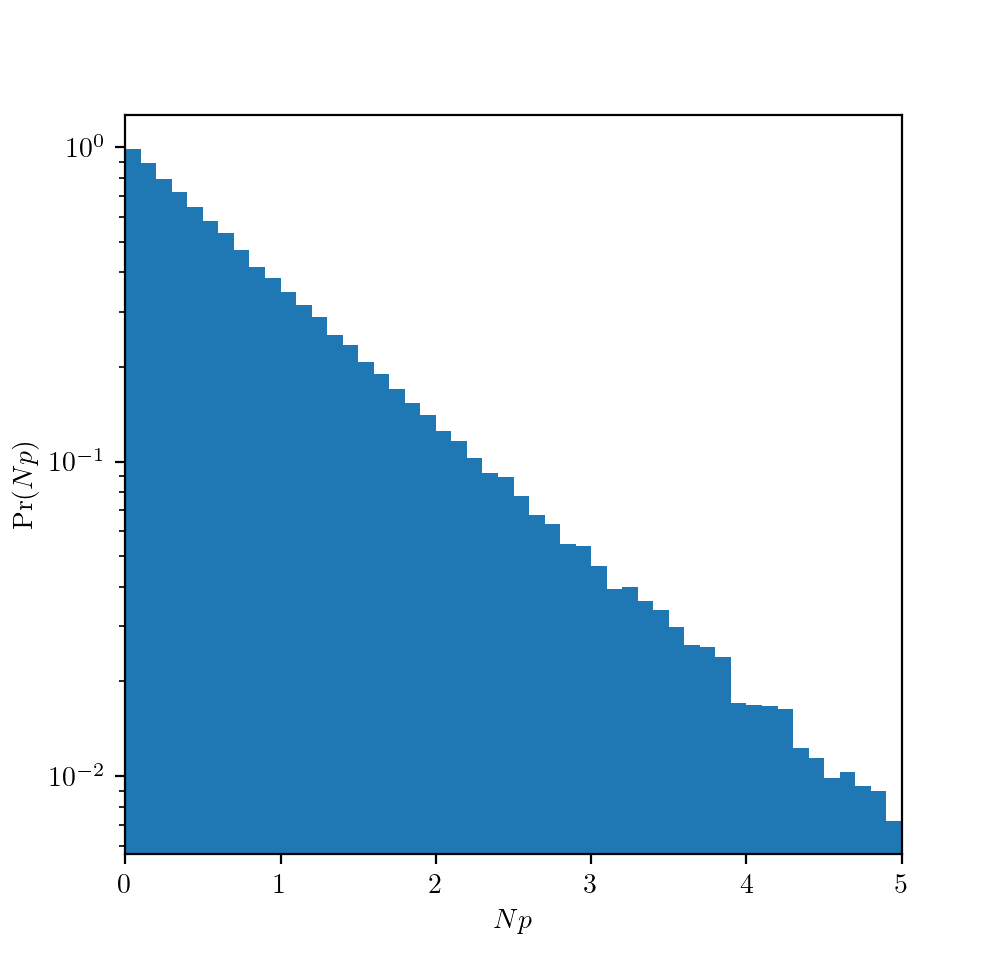}
  \caption{The distribution of measurement probabilities from the 200,000 amplitudes calculated for Bristlecone-70 circuits with depth 1+28+1 . It closely matches the Porter-Thomas form $Ne^{-Np}$.}
  \label{fig:pt}
\end{figure}

\textbf{Simulating Bristlecone-70 circuits with AC-QDP v1.1}: A large-scale simulation of \\Bristlecone-70 circuits using double precision was conducted on an Alibaba Cloud cluster composed of $1449$ Elastic Compute Service (ECS) instances, each with $88$ Intel Xeon (Skylake) Platinum 8163 vCPU cores @ 2.5 GHz and 160 GiB of memory as computation nodes. We calculated $200,000$, $1,000$, $200$, and $1$ amplitudes of Bristlecone-70 circuits with depth $1+28+1$, $1+32+1$, $1+36+1$, and $1+40+1$, respectively. The last two were the first successful simulations of instances at those depths. For the sake of comparison, we also created $4$ ECS instances with $2$, $4$, $8$, and $16$ vCPU cores, respectively, as part of a `virtual' cluster of $127, 512$ vCPU cores. Those ECS instances supported Memory-to-CPU-core ratios of $2$. For a large tensor network contraction task that has been split into $K$ subtasks that each fit into $2$ GiB of memory, we ran $\frac{K}{127,512}$ subtasks on each vCPU core. We use the largest execution time on those ECS instances to extrapolate the execution time of the whole tensor network contraction task on the large cluster. Comparing the results, we concluded that scheduling and communication costs are negligible. Detailed benchmarking results can be found in Table~\ref{tab:bristlecone} and \cite{ZHN+19}. Based on the 200,000 amplitudes we calculated for Bristlecone-70 circuits of depth 1 + 28 + 1, we plot the distribution of $N_p$ in Figure~\ref{fig:pt}, which closely matches the Porter-Thomas form.

\begin{table}[htb!]
  \begin{tabular}{ |p{5.1cm}||p{3cm}|p{3cm}|p{3cm}|  }
    \hline
    \multicolumn{4}{|c|}{Extrapolated execution time on a cluster with $127,512$ vCPU cores} \\
    \hline
    Circuit & Amplitudes & Subtasks p.a.& Seconds p.a.\\
    \hline
    Bristlecone-70$\times$(1 + 28 + 1)   & 200,000    & 256 &   0.03\\
    Bristlecone-70$\times$(1 + 32 + 1)   & 10,000  & 1024   & 0.36\\
    Bristlecone-70$\times$(1 + 36 + 1)   & 100 & 65536&  4.56\\
    Bristlecone-70$\times$(1 + 40 + 1)   & 1 & 4194304&  480.17\\
    \hline
    \hline
    \multicolumn{4}{|c|}{Actual execution time on a cluster with $127,512$ vCPU cores} \\
    \hline
    Circuit & Amplitudes & Subtasks p.a.& Seconds p.a.\\
    \hline
    Bristlecone-70$\times$(1 + 28 + 1)   & 200,000    &256&   0.04\\
    Bristlecone-70$\times$(1 + 32 + 1)   & 1,000  & 1024   &0.43\\
    Bristlecone-70$\times$(1 + 36 + 1)   & 200 & 65536&  5.75\\
    Bristlecone-70$\times$(1 + 40 + 1)   & 1 & 4194304&  580.7\\
    \hline
  \end{tabular}
\caption{\label{tab:bristlecone} Extrapolated and actual execution times for the simulation of random circuit sampling on Bristlecone by using AC-QDP version 1.1 on an Alibaba Cloud cluster of $127,512$ vCPU cores.}
\end{table}

\textbf{Simulating Sycamore-53 circuits  with AC-QDP v2}: We use a `virtual' Alibaba Cloud cluster, with resources comparable to the Summit supercomputer, to estimate the cost of the proposed supremacy task, namely sampling one instance of a given quantum circuit a million times with a given XEB fidelity. Our `virtual' cluster consists of $27,648$ \textit{ecs.gn6v-c8g1.2xlarge} instances acting as computational nodes and an \textit{ecs.g6.6xlarge} instance acting as an agent node to handle preprocessing and scheduling. Since our previous large-scale simulation demonstrated that scheduling and communication times are negligible, we can use the execution time on one computational node to extrapolate the execution time on the cluster. The agent node, an \textit{ecs.g6.6xlarge} instance, has $24$ Intel Xeon (Skylake) Platinum 8163 vCPU cores @ 2.5 GHz and $96$ GiB of memory. A computation node, an \textit{ecs.gn6v-c8g1.2xlarge} instance, has $8$ Intel Xeon (Skylake) Platinum 8163 vCPU cores @ 2.5 GHz and $32$ GiB of memory, as well as an Nvidia Tesla V100 SMX2 GPU with $16$ GiB of RAM. We use single precision. In order to make best use of the GPU, we use the JAX library to just-in-time compile our contraction scheme. Since all subtasks for a sampling task use the same contraction scheme, the compilation only needs to be done once on each computation node, and the overall time spent is negligible. Detailed benchmarking results can be found in Table~\ref{tab:sycamore}.

\subsubsection{Comparison with other simulators}
We performed extensive comparisons with other tensor network-based simulators, including qFlex\cite{VBN+19,AAB+19} and Cotengra\cite{GK20}. 

In \cite{AAB+19}, benchmarked runtimes and FLOPs of calculating batches of 64 amplitudes on the Summit supercomputer at Oak Ridge National Laboratory were reported. We perform our runtime estimation by using exactly the same number of Elastic Compute Service nodes (27648) with exactly the same graphics card model (Nvidia Tesla V100 SXM2 16 GB) as the Summit supercomputer, but on Alibaba Cloud. We tested our framework across different Alibaba Cloud regions, which could affect the performance due to lack of low-latency interconnects across regions. However, because our algorithm introduces very little communication, the communication overhead is negligible.

In \cite{GK20}, benchmarked times and other information for computing single amplitudes of random circuits with single precision using Nvidia Quadro P2000s were reported. In order to conduct a fair comparison, we re-ran the latest GitHub version of Cotengra to calculate single amplitudes with perfect fidelity on the same Elastic Compute Service node with the same graphics card (Nvidia Tesla V100 SXM2 16 GB) that we used to benchmark AC-QDP. In AC-QDP, the preprocessing step is controlled by the number of iterations in the CMA-ES algorithm and the number of iterations of local optimization.  Consequently, there is no way to precisely control the preprocessing time. However, by using the iteration parameters mentioned above, it usually takes from one hour to several hours, depending on number of cycles. To avoid underestimating Cotengra and to better understand its ultimate capability, we allow the path optimizer in Cotengra to search for 10 hours, which should always be large enough to compare with the pre-processing time used in the AC-QDP benchmark.
As contracting open tensor networks has not been implemented in the latest version of Cotengra on GitHub (949635c6783435fc384553ec28c7c038dc786e01), we use the runtime reported for a single amplitude calculation as a lower bound for a 64-amplitude batch calculation. We then use the runtime of a single-amplitude calculation to estimate the cost of frugal rejection sampling with $10\times$ overhead as an upper bound for the sampling task  \cite{MFI+18}.  

We also compare AC-QDP with the classical algorithms proposed in \cite{AAB+19}. Since both are targeting the same sampling task by assuming Summit-comparable computational resources, we use the numbers reported in \cite{AAB+19}.

The simulation proposal in \cite{PGN+19} leverages secondary storage to assist main memory when the quantum state is too large to fit. Any proposal that needs to store the entire state vector is limited by the storage space available. The Summit supercomputer has 250 PiB of secondary storage, while a 54-qubit state vector stored in single precision takes 128 PiB to store. The space requirement doubles with each additional qubit, and so this proposal would already have trouble scaling to 55-qubit circuits, while 56-qubit circuits would be out of reach.

The time estimate in \cite{PGN+19} is based on the runtimes reported in \cite{HS17}, in which $0.5$ PiB of memory were used to support a double-precision simulation of a $45$-qubit quantum circuit on the Cori II supercomputer. The simulation makes use of highly optimized CPU kernels, which are not applicable to Summit, a supercomputer with most of its computational power in GPUs. Furthermore, \cite{PGN+19} scales the runtime down based on LINPACK benchmark figures for Cori II and Summit, making two implicit assumptions.  First, the CPU kernels in \cite{HS17} that are highly optimized for the task of quantum simulation are portable to GPUs with no loss in efficiency. Second, the LINPACK benchmark figures, which allow the user to choose the most suitable problem size for a given machine, are applicable to the specific task that \cite{HS17} performs on a single node: simulating up to 32 local qubits. This is problematic given that each socket (the Summit equivalent of a node in \cite{PGN+19}) only has $16\times3=48$ GiB of GPU memory among its 3 GPUs, while data transfer to and from the GPU can be quite expensive.

For comparison, we estimated the theoretical runtime of our simulator for the 20-cycle quantum supremacy task under two analogous assumptions. First, a further optimized implementation should achieve a GPU FLOPS efficiency comparable to Cotengra, a simulator fundamentally similar to ours. In 18- and 20-cycle experiments on V100 GPUs, we observe a GPU efficiency in excess of $90\%$ with respect to the single-precision performance figure of 15.7 teraFLOPS. Second, we assume that tensors of rank 32 can be contracted on GPUs, presumably using out-of-core memory access, without a significant loss in efficiency. We would then only need to slice enough indices to decrease the contraction width to 32, for which we found a contraction scheme with total cost $10^{18.6}$. Combining both assumptions yields the estimate
$$\frac{(8 \times 10^{18.6})\textrm{ FLOPs/sample}}{15.7\times 10^{12} \textrm{ FLOPS}\times 90\%\times 27648}\times 2000\textrm{ samples} \approx 1.63\times 10^5\text{ seconds} \approx 1.89\text{ days}.$$

\end{document}